\documentclass[aps,prb]{revtex4} 
\usepackage{graphicx}
\usepackage{epsfig}

\begin{document}
\title{Optimal prediction of folding rates and transition
state placement from native state geometry}
\author{Cristian Micheletti}
\affiliation{International School for Advanced Studies (S.I.S.S.A.) and
INFM, Via Beirut 2-4, 34014 Trieste, Italy}
\date{\today}
\begin{abstract}              
A variety of experimental and theoretical studies have established
that the folding process of monomeric proteins is strongly influenced
by the topology of the native state. In particular, folding times have
been shown to correlate well with the contact order, a measure of
contact locality. Our investigation focuses on identifying additional
topologic properties that correlate with experimentally measurable
quantities, such as folding rates and transition state placement, for
both two- and three-state folders. The validation against data from
forty experiments shows that a particular topologic property which
measures the interdepedence of contacts, termed cliquishness or
clustering coefficient, can account with significant accuracy both for
the transition state placement and especially for folding rates, the
linear correlation coefficient being $r=0.71$. This result can be
further improved to $r=0.74$, by optimally combining the distinct
topologic information captured by cliquishness and contact order.
\end{abstract}

\maketitle

\section{Introduction}

In the past three decades, there has been a growing effort of the
scientific community for studying and understanding the principles
that govern the folding process of a sequence of amino acids in the
corresponding native structure \cite{anfinsen,creighton,tooze}.  In
recent years, several proteins, in particular those folding via a
two-state mechanism \cite{1coa} have provided an extraordinary
benchmark for experimental and theoretical characterization of the
folding pathways. The significant amount of experimental data
available for several structurally unrelated proteins
\cite{1coa,Jackson,lmb,lmb2,2abd,2abd2,1ycc,2ait,1mjc,1srl,1pba,1ubq,2ptl,1urn,1hdn,1aps,1imq,2gb1,1div,1fnf,1cis},
has opened the possibility to identify and isolate the factors that
influence the folding rate. Besides considering detailed chemical
interaction, such as those affecting free-energy barriers, an
appealing and elegant line of investigation has focused on the effects
of the native state structure on the folding process
\cite{Plaxco,goddard,karplusnsb,Finkelstein01}.

From a qualitative point of view, the influences of structural effects
was traditionally summarised in the tenet that proteins with high
helical content fold faster than proteins with mixed alpha/beta
content, the slowest folding being for the all-beta ones. This useful
and intuitive rule of thumb, fails to account for the very different
rates observed between proteins in each of the alpha, alpha/beta or
beta families \cite{Jackson,aurora,M00,capaldi}. A deep insight into
this problem was provided by the work of Plaxco {\em et al.}
\cite{Plaxco} who introduced the concept of contact order, which
captures, quantitatively, features beyond the mere secondary structure
motifs. The highly significant correlation of contact order and
experimental folding rates shows the extent to which the mere topology
of native state can influence the folding process. However, the highly
organised native structure of proteins is too rich to be captured by a
single parameter such as the contact order. Indeed, the latter cannot
account in the same satisfactory way for the transition state
placement, three-state folding rates or the diversity of folding rates
among structurally similar proteins \cite{Igfolds}.

In the present study we investigate how the topology of the native
state can be further exploited to provide optimised predictions for
protein folding rates and the transition state placement. To do so we
consider, among others, one particular topological descriptor that is
crucial for characterising the connection and interactions of native
contacts: the clustering coefficient , or cliquishness. Such
parameter, heavily studied in the context of graph theory
\cite{bollabas,strogatz,strogatz2} is shown to have highly significant
correlation with folding rates. The advantage of using this topologic
descriptor is that it allows to capture the cooperative formation of
native interactions, as proved by its statistically relevant
correlations with the transition state placement. Further, we discuss
how the different topologic aspects captured by the cliquishness and
contact order can be combined to yield optimal correlations higher
than for the individual descriptors.

\section{Theory and Results}

Customarily, at the heart of theoretical or numerical studies of
topology-based folding models is the contact matrix (or map) \cite{Go}
which will be used extensively also in the present context. The
generic entry of the contact map, $\Delta_{ij}$, takes on the value 1
if residues $i$ and $j$ are in contact and zero otherwise. Several
criteria can be adopted to define a contact; in the present study we
consider two amino acids in interaction if any pair of heavy atoms in
the two amino acids are at a distance below a certain cutoff, $d$. All
values of $d$ between 3.5 \AA\ and 8 \AA\ have been considered and
reported.  The contact map provides a representation for the spatial
distribution of contacts in the native structures that is both concise
and often reversible (since native structures can be recovered when
appropriate values of $d$ are used). Plaxco and coworkers\cite{Plaxco}
have used the contact map to describe and characterize the presence
and organization of secondary motifs in protein structures. The
parameter that was introduced, the relative contact order, provides a
measure of the average sequence separation of contacting residues and
is defined as

\begin{equation}
\mbox{relative contact order} = {1 \over L} {\sum_{i \not= j} \Delta_{ij}\ | i
-j|\ w_{ij}
\over \sum_{i \not= j} \Delta_{ij}\ w_{ij}},
\label{eqn:co}
\end{equation}

\noindent where $i$ and $j$ run over the sequence indeces, $w_{ij}$ is
the contact degeneracy (i.e. the number of pairs of heavy atoms in
interaction) and $L$ is the protein length. Remarkably, the contact
order was shown have a highly significant linear correlation with
experimental folding rates. The result of Plaxco and coworkers can be
explained, {\em a posteriori}, with intuitive arguments: a high
contact order corresponds to few local interactions. One may thus
expect that the route from the unfolded ensemble to the native state
is slow, being hindered by the overcoming of several barriers
\cite{Funnel1,Funnel2,Funnel3,Funnel4,Funnel5} due to spacial
restraints, as recently analysed by Debe and Goddard \cite{goddard}
and previously by Chan and Dill \cite{chan90} and also observed in
topology-based numeric studies \cite{hiv}.  These considerations are
based purely on geometric arguments and do not take into account the
influence of specific interactions between the residues. In principle,
the latter may well override the topological influence on the folding
process, but surprisingly, as remarked in a recent review article
\cite{bakernature} this is often not the case
\cite{hiv,Miche99b,Clem,Finkel,eaton,baker,thirum}.

Our aim is to exploit as much as possible the topologic information
contained in the native state to improve both the accuracy of
predictions for folding rates and gain more fundamental insight into
the process. To this purpose we have considered additional topologic
descriptors besides the contact order. The one that appeared most
significant is a parameter termed cliquishness or clustering
coefficient \cite{bollabas,strogatz,strogatz2}. For a given site, $i$,
the cliquishness is defined as:

\begin{equation}
\mbox{cliquishness}(i) = {\sum_{j \not =l} \Delta_{ij} \Delta_{il}
\Delta_{lj} \over \sum_{j \not =l} \Delta_{ij} \Delta_{il}} = {\sum_{j \not =l} \Delta_{ij}
\Delta_{il} \Delta_{lj} \over N_c(N_c -1)/2} ,
\label{eqn:cliq}
\end{equation}

\noindent where $N_c$ is the number of contacts to which site $i$
takes part to. As for the contact order, also the cliquishness has an
intuitive meaning; in fact it provides a measure of the extent to
which different sites interacting with $i$ are also interacting with
each other. Of course, the cliquishness is properly defined only if
site $i$ is connected to, at least, two other sites. To ensure this,
we included also the covalently bonded interactions $[i,i\pm 1]$ in
(\ref{eqn:cliq}). The importance of taking the cliquishness into
account for discriminating fast/slow folders can be anticipated since
a higher interdependency of contacts (large cliquishness) will likely
result in a more cooperative folding process. In fact, the formation
of a fraction of interactions will result in the establishment of a
whole network of them. Consistently with this intuitive picture one
should also expect that a large/small cliquishness will affect in
different ways the amount of native-like content of the transition
state.

We have tested and verified these expectations by calculating the
average cliquishness for 40 proteins for which folding rates and
transition state placement, $\theta_m$, have been measured. $\theta_m$
is deduced from the variation of folding/refolding rates upon change
of denaturant concentration ($m^{\dagger}_F$ and $m ^\dagger_U$) and
provides an indirect indication of how much the solvent-exposed
surface of the transition state is similar to that of the native
one. $\theta_m$ ranges between 0 and 1; higher values denote stronger
similarity with the native state. It is worth pointing out that,
although the model underlying the calculation of $\theta_m$ relies on
a two-state analysis, an effective $\theta_m$ can be inferred for
three-state folders as well \cite{Jackson}.  Since reliable
$\theta_m$'s are not available for all proteins, the number of entries
used to correlate the cliquishness and $\theta_m$ (see Tables
\ref{tab:list1} and \ref{tab:list2}) is slightly smaller than that
used for tge logarigthm of refolding rates, $\ln K_F$. The set of
proteins used, shown in Tables \ref{tab:list1} and \ref{tab:list2},
was built up from experimental data collected in previous studies and
predictions (often topology-based) of folding rates
\cite{Jackson,Plaxco,goddard,karplusnsb,Finkelstein01}.

As indicated, the entries include both two-state and three-state
folders, proteins belonging to the same structural family as well as
proteins under different experimental conditions.  This allows to
examine to what extent predicted folding rates are consistent with the
wide variations of folding velocities observed in structurally-related
proteins and in different experimental conditions. As discussed in
detail below, when the comprehensive set of Table \ref{tab:list1} and
\ref{tab:list2} is used, the correlation found between cliquishness
and folding rates is 0.71, with a statistical significance of $t=
10^{-5}$, more relevant than the one between a suitably defined contact
order and folding velocities ($r=0.66$, $t= 5 \cdot 10^{-5}$). As will
be shown, the predicting power of the two quantities can be combined
to achieve the optimal correlation of 0.74.
 
The prediction of the transition state placement, turns out to be more
difficult when either of the two topologic parameters is used. While
for the contact order it is equal to 0.23, the cliquishness yields the
value of 0.48 which is not significantly improved by combining the two
descriptors. Though the linear correlation of the clustering
coefficient and the transition state placement is not as high as for
the folding-rate case it is nevertheless statistically meaningful,
having a probability of 0.004 to have arisen by chance.

\subsection{Two- and three-state folders}

Before considering the more general case of all entries in Tables
\ref{tab:list1} and \ref{tab:list2}, we focus on two-state folders,
i.e.  proteins with a cooperative (all-or-none) transition between the
unfolded and folded states. The neatness of this process, due to the
absence of any significantly populated intermediate state, makes them
ideal candidates for identifying and isolating the factors that
influence the folding rate.

In the present context this separate test is important since it
appears that the relative contact order is a much stronger descriptor
for two-state folders, than for the general case. As a matter of fact,
when both two- and three-state folders are considered, the influence
of the average sequence separation of native contacts on folding
properties is better captured by a different version of the contact
order, which we shall term ``absolute'', obtained when the r.h.s. of
eq. (\ref{eqn:co}) is not divided by $L$:

\begin{equation}
\mbox{absolute contact order} = {\sum_{i\not= j} \Delta_{ij}\ | i -j|\ w_{ij}
\over \sum_{i \not= j} \Delta_{ij}\ w_{ij}}\ .
\label{eqn:coa}
\end{equation}

\noindent In the following we shall report and compare the performance
of both parameters; furthermore we shall always consider the absolute
value of the linear correlation coefficients, $|r|$, without regard to
its sign, which can be easily inferred from the plots.

\begin{figure}
\begin{center}
\epsfig{figure=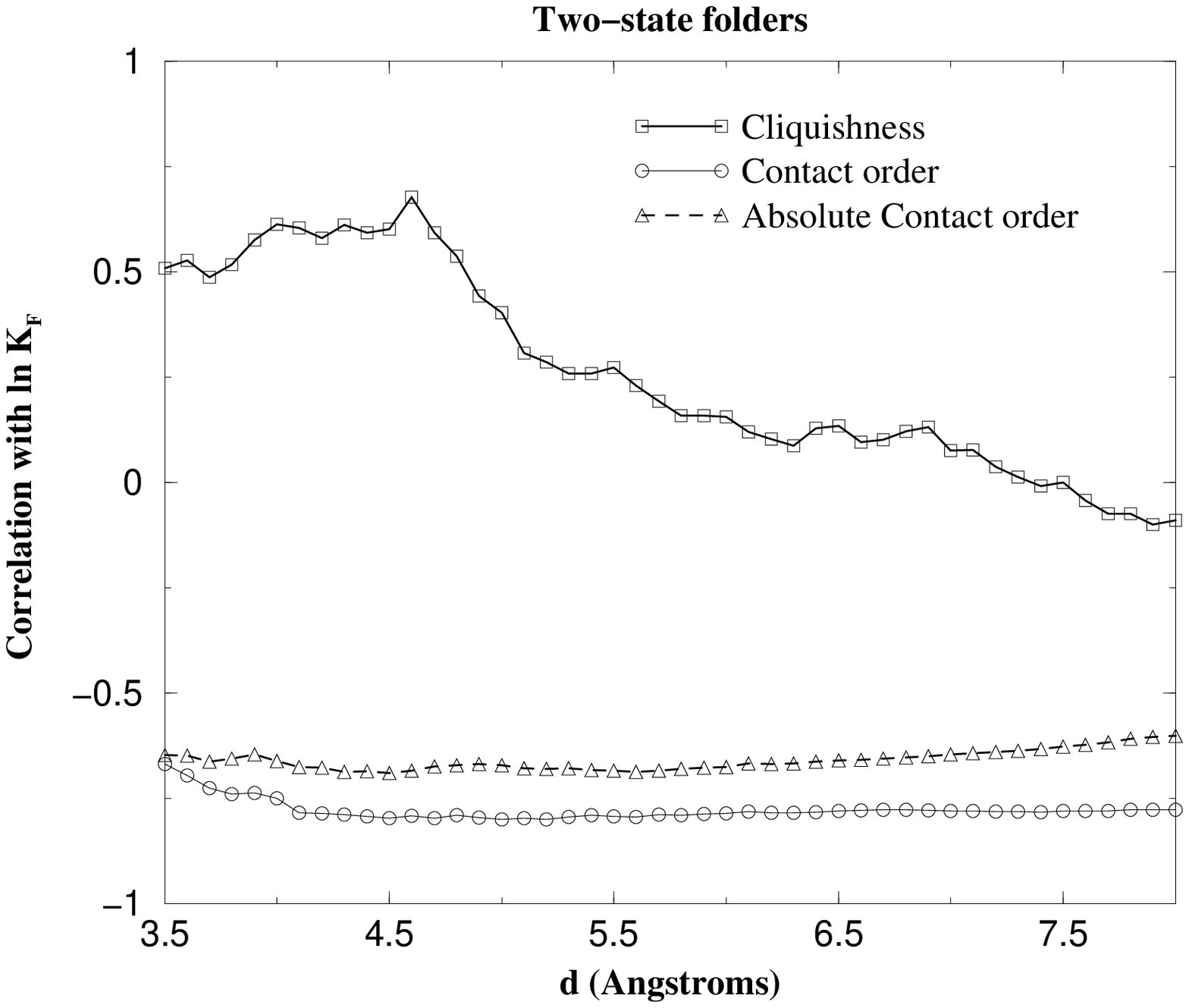,width=0.80\textwidth}
\end{center}
\caption{}
\label{fig:2fr}
\end{figure}

The original definition of contact order has an unrivaled performance
in the prediction of folding rates for the two-state folders of Table
\ref{tab:list1}. As visible in Fig. \ref{fig:2fr}, it gives a stable
correlation for cutoffs in the range 5 \AA $ \le d \le$ 7 \AA, with
the maximum value of $r=0.80$ for the cutoff $d=4.5$ \AA. The
statistical significance of such correlation can be quantified through
a calculation of the probability, $t$, to observe by pure chance a
correlation higher than the measured one (in modulus). The standard
model underlying such estimates relies on the hypothesys of normal
distribution of the deviates of the correlated quantities. As a rule
of thumb, the upper value of $t=0.05$ is taken as a threshold for
statistically meaningful correlations.  For the value of $r=0.80$
reported above, this probability is $t=3 \cdot 10^{-5}$, which is,
therefore, extremely significant.

\begin{figure}
\begin{center}
\epsfig{figure=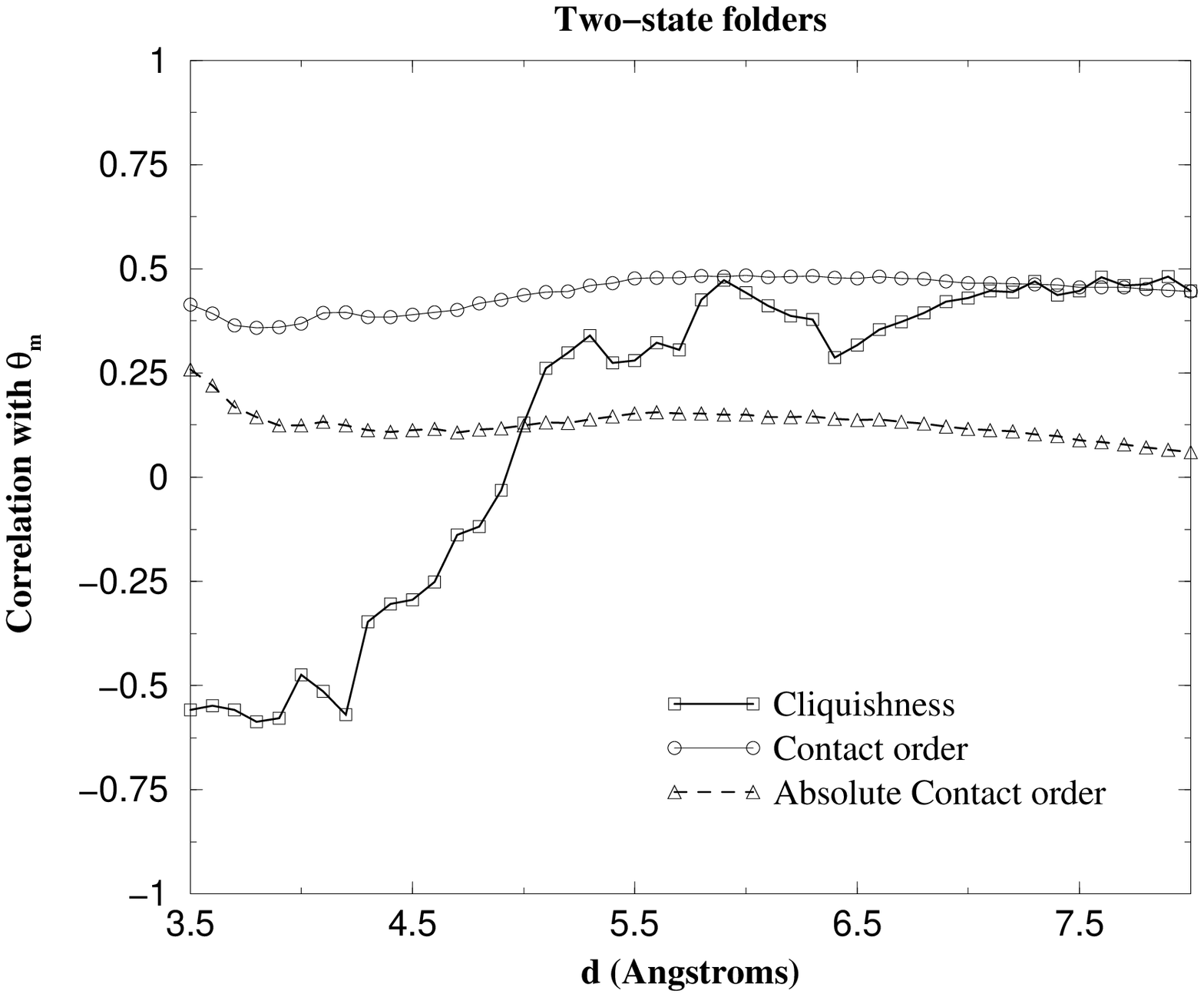,width=0.80\textwidth}
\end{center}
\caption{}
\label{fig:2theta}
\end{figure}

Consistently with previous results, we found that the transition state
placement is a much more elusive quantity to predict than folding
rates. In fact, all topologic descriptors yield a poorer correlation
compared to $\ln K_F$ (see Fig. \ref{fig:2theta}). For the relative
contact order, the best $r$ is 0.48 (for $d=6.0$) with an associated
$t=0.02$. As anticipated, the performance of the absolute contact
order in this particular context is significantly inferior then the
relative one (see Figs. \ref{fig:2fr} and \ref{fig:2theta}) and hence
will not be further commented.

Concerning the performance of the novel parameter under scrutiny, the
cliquishness, it can be seen from Figs.  \ref{fig:2fr} and
\ref{fig:2theta} that it is statistically meaningful for both folding
rates and transition state placement. There are, however, significant
differences with respect to the contact order. For folding rates the
optimal $r$ is 0.67 ($d=4.6$ \AA) and the associates value of $t$ is 5
$10 ^ {-4}$, one order of magnitude larger than for the relative
contact order. For $\theta_m$ the situation is reversed since the
optimal value of $r=0.58$ (for $d=3.8$ \AA) has the statistical
relevance of $t=0.004$, with a marked improvement over the
previous case.

It is also interesting to note that cliquishness-based correlations
have a non-trivial dependence on the cutoff $d$. In fact, due to the
overall compactness and steric effects, the degree of dispersion of
the cliquishness values for different sites in the same or different
proteins is much more limited compared, e.g. to the average sequence
separation of contacts. This leads to the observed decay of the
correlations when the cutoff $d$ is increased.

\begin{figure}
\begin{center}
\epsfig{figure=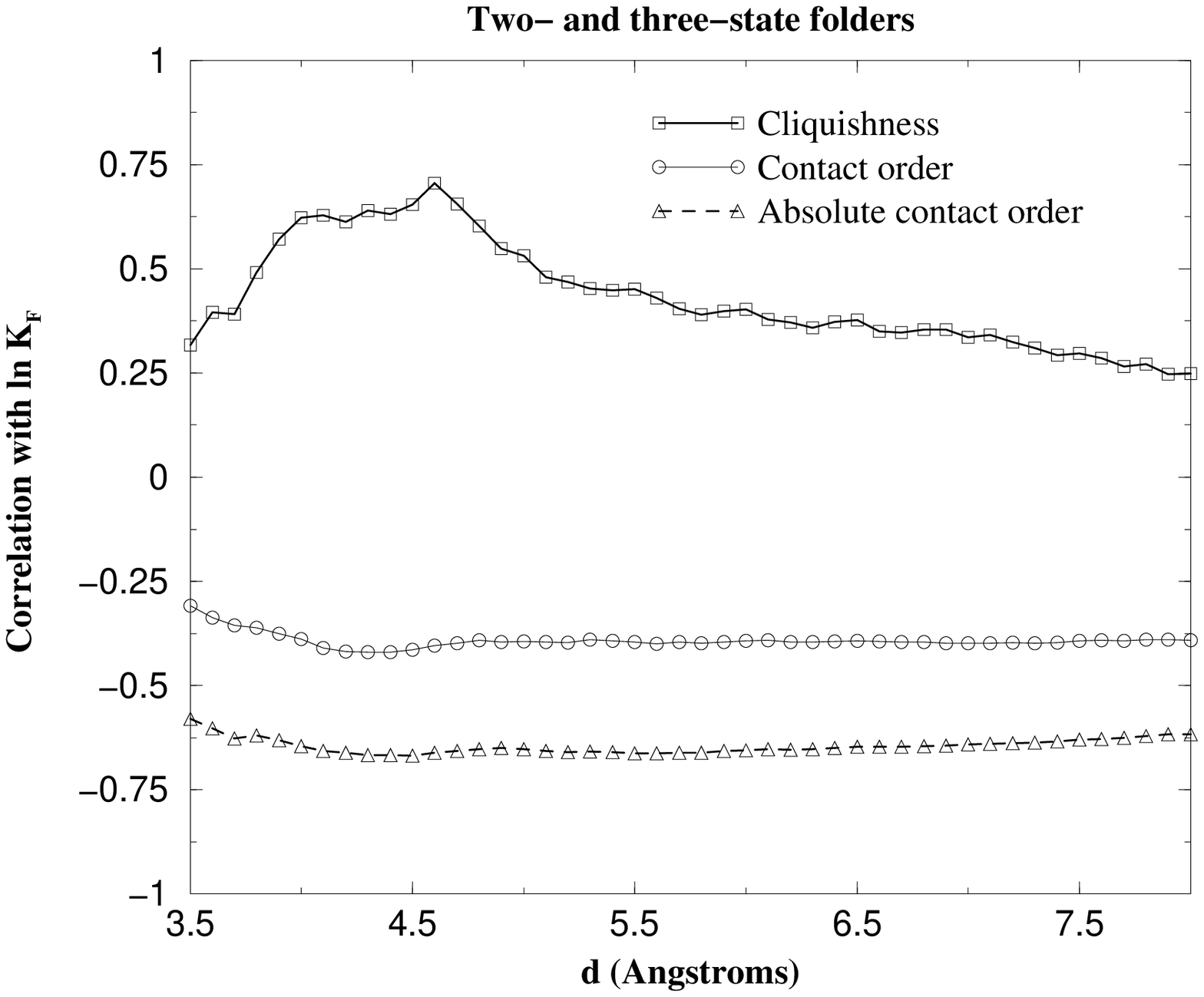,width=0.80\textwidth}
\end{center}
\caption{}
\label{fig:2_3fr}
\end{figure}

\begin{figure}
\begin{center}
\epsfig{figure=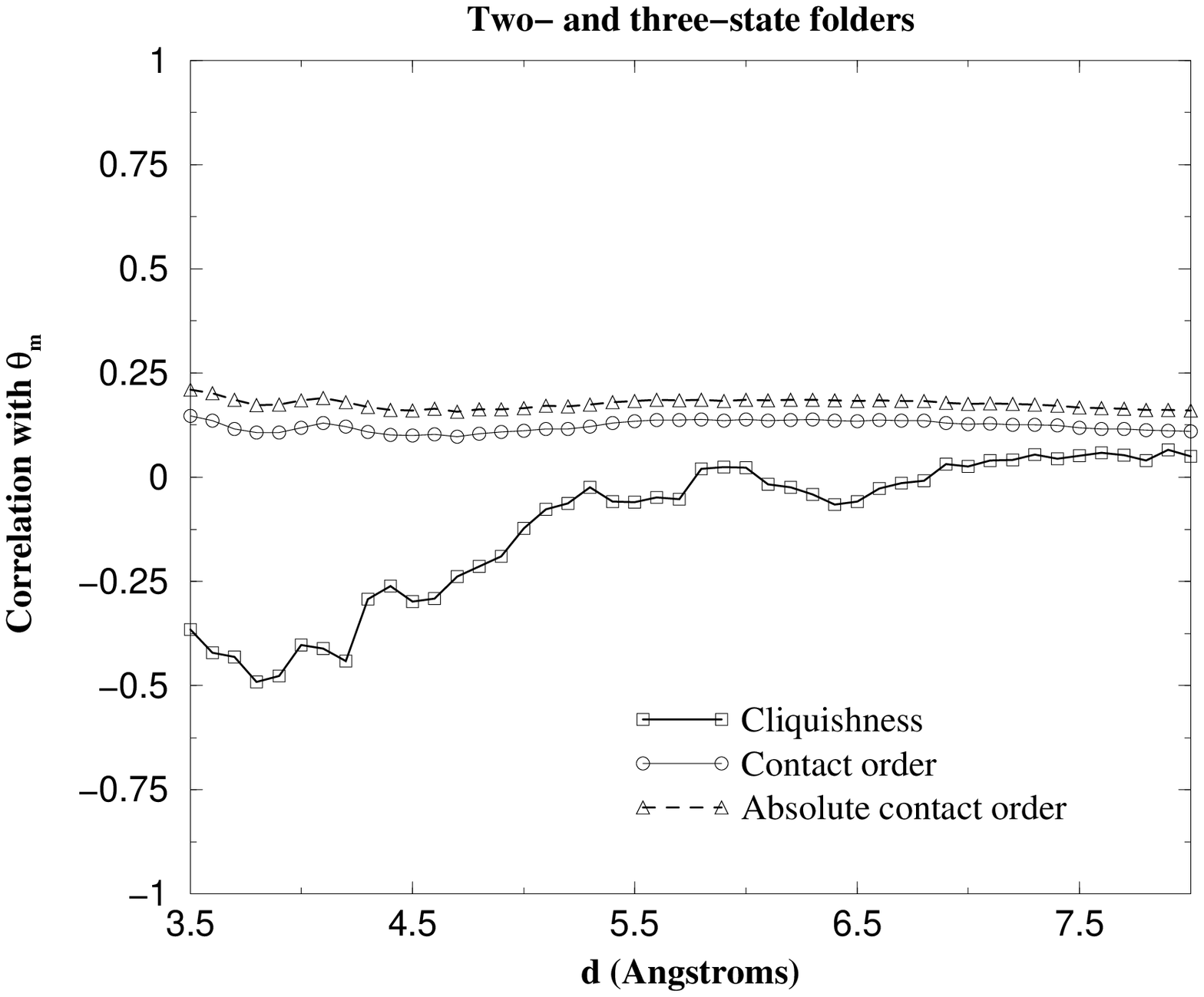,width=0.80\textwidth}
\end{center}
\caption{}
\label{fig:2_3theta}
\end{figure}

The applicability of topology-based models are not limited to
two-state folder, but can be extended to include three-state folders
as well \cite{Miche99b,Clem}. Despite the addition of the 11 entries
corresponding to three-state folders, the performance of
cliquishness-based predictions for folding rates and $\theta_m$
improves from the values reported for two-state folders. As shown in
Figs \ref{fig:2_3fr} and \ref{fig:2_3theta} the associated optimal
correlations for $\ln K_F$ and $\theta_m$ are $r=0.71$ and 0.49, again
observed for the same cutoff values ($d$) mentioned for the two-state
case. The corresponding statistical significances are now, $t=1\cdot 10
^{-5}$ and $t=0.004$, which, despite the enlargement of the
experimental set, show even an improvement over the two-state case.

From Figs. \ref{fig:2_3fr} and \ref{fig:2_3theta} it can be noticed
that the performance of the relative contact order is noticeably
poorer than the absolute contact order which, being a much better
descriptor, becomes the focus of our analysis. The corresponding
measured correlations are, in fact, $r=0.66$ for $\ln K_F$ and
$r=0.20$ for $\theta_m$ with corresponding values of $t=5\cdot 10 ^{-5}$
and $t=0.23$.

\begin{figure}
\begin{center}
\epsfig{figure=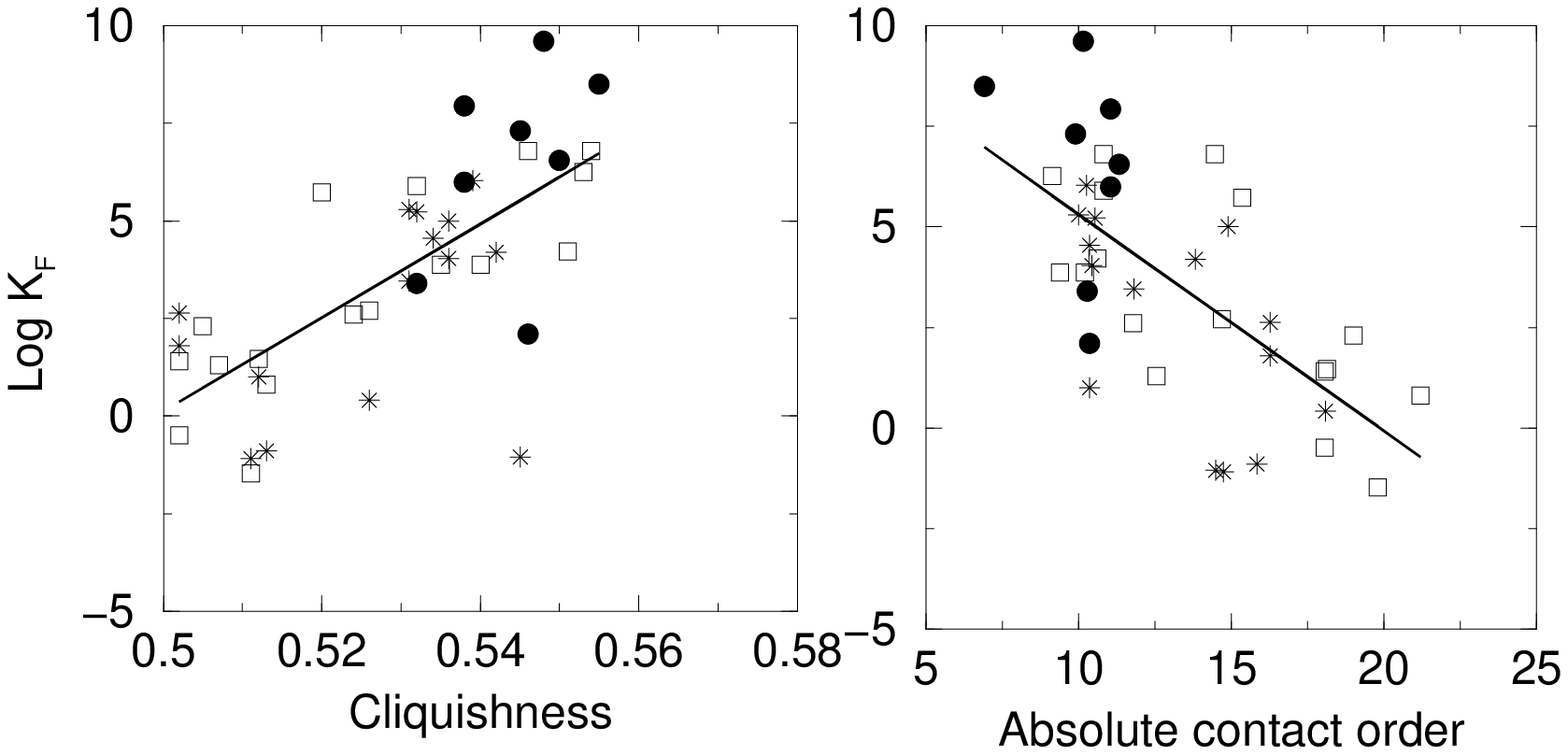,width=0.80\textwidth}
\end{center}
\caption{}
\label{fig:best_fr}
\end{figure}

\begin{figure}
\begin{center}
\epsfig{figure=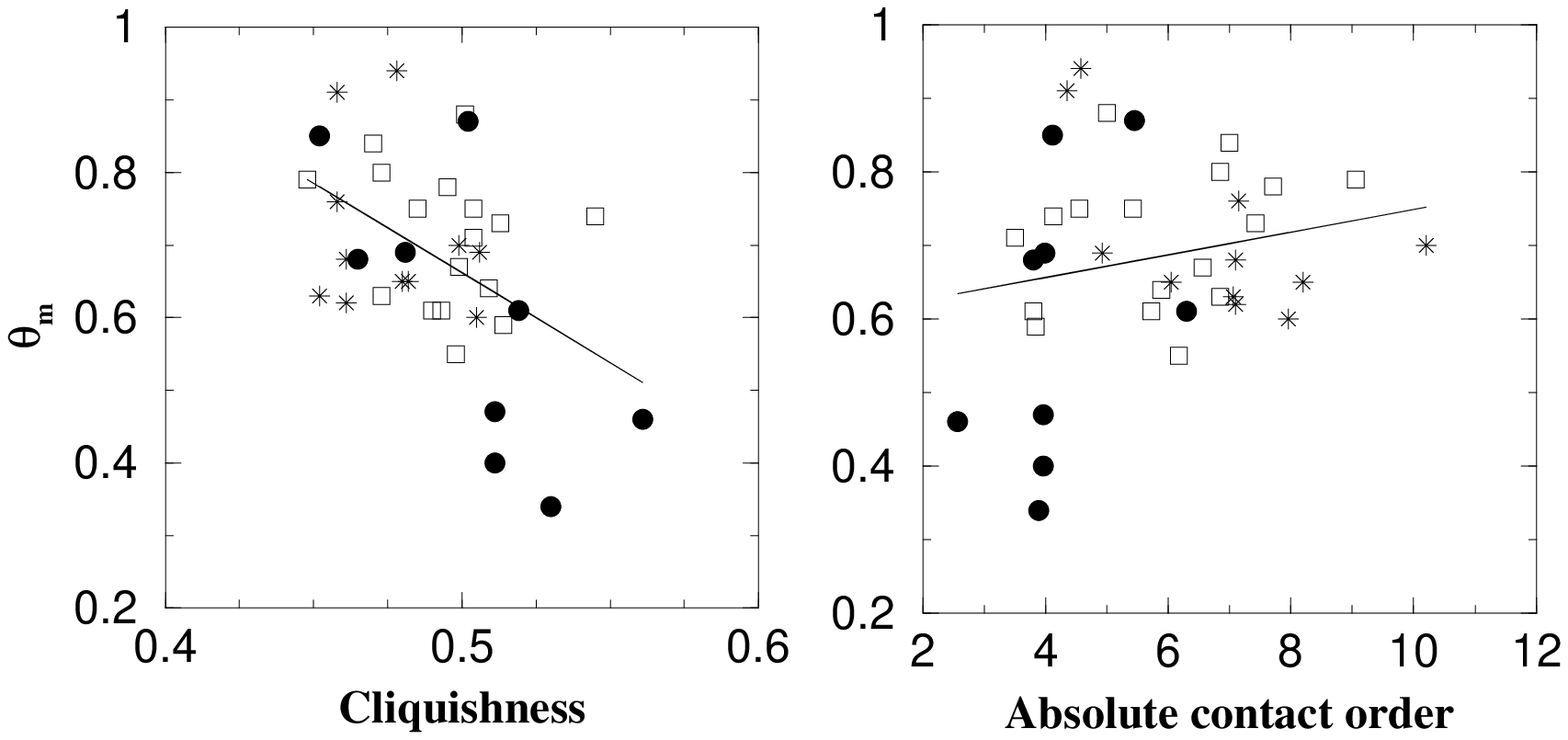,width=0.80\textwidth}
\end{center}
\caption{}
\label{fig:best_theta}
\end{figure}

A direct comparison of how the clustering coefficient and the absolute
contact order correlate with $\ln K_F$ and $\theta_m$ can be made by
inspecting the plots of Figs. \ref{fig:best_fr} and
\ref{fig:best_theta}. It is worth pointing out that the analysis of
the deviations from the linear trends of Figs. \ref{fig:best_fr} and
\ref{fig:best_theta} reveals that a particular protein, 1urn, is among
the top outliers for both cliquishness and contact order-based
analysis, although no simple explanations is available for this
singular behaviour.  Although for both folding parameters the
cliquishness gives a more significant correlation than contact order,
the difference is particularly dramatic for the transition state
placement which is notoriously difficult to capture with
topology-based predictions \cite{Plaxco}.

An important conclusion stemming out of this observation is that the
transition state structure (and hence $\theta_m$) is more influenced
by the degree of interdependency of native contacts than their average
sequence separation. This is in accord with the intuition that highly
interdependent contacts may mutually enhance their probability of
formation, thus facilitating the progress towards the native state
during the folding process. This is, indeed, consistent with the
negative correlation observed between cliquishness and native content,
$\theta_m$, at the transition state. It is important to stress that
the presence and effects of the cooperative formation of native
interactions cannot be captured by parameters based on measures of
contact locality. This highlights the importance of considering all
viable topologic descriptors to characterize the folding process,
since they do not impact in the same way on various folding
properties.

\subsection{Optimal combined correlation}

A natural question that arises is whether it is possible to combine
the predicting power of cliquishness and contact order to achieve
correlations with experimental folding rates and transition state
placements that are better than the individual cases.

Indeed, as shown in Appendix A, it is straightforward to combine in an
optimal linear way the two quantities to improve the prediction
accuracy. The quantitative increment in the correlation is clearly
related to the amount of independent information contained in the two
topologic descriptors. Hence, an important issue is to what extent
cliquishness and contact order are mutually correlated.

If, in place of a physical contact map, $\Delta_{ij}$, one uses a
random symmetric matrix, no meaningful correlation will be found.  The
contact maps of real proteins, however, display features that are
highly non-random which reflect both (i) the physical constraints to
which a compact three-dimensional chain is subject and (ii) the
presence and organisation of secondary motifs
\cite{Chothia2,Chothia1,Levitt,Rose}.

With the aid of numeric simulations it was possible to assess the
degree of interdependency of clustering coefficient of native contacts
and their average sequence separation resulting from the first of the
mentioned effects. This was accomplished by considering, in place of
the proteins of tables \ref{tab:list1} and \ref{tab:list2}, 150
computer-generated compact structures respecting basic steric
constraints found in real proteins (details can be found in the
Methods section). As visible in the plot of Fig. \ref{fig:decoys} the
level of mutual contact order-cliquishness correlation observed in
these artificial structures is $r=0.25$ which is significantly smaller
than the actual correlation of the two quantities found in real
proteins. In fact, the typical correlation for cliquishness and
contact order (either relative or absolute) is around 0.65.  Such non
trivial correlation can been ascribed to the special topologic
properties of naturally occurring proteins whose ramifications have
been investigated in a variety of contexts
\cite{M00,opthelix,plato,Miche99b,Hunt94,Socci94}. Thus, the very
presence and organization of secondary motifs in proteins makes it
possible, on one hand, to exploit the native topology to predict e.g.
folding rates, while on the other it limits the amount of independent
information contained in different topologic descriptors.

\begin{figure}
\begin{center}
\epsfig{figure=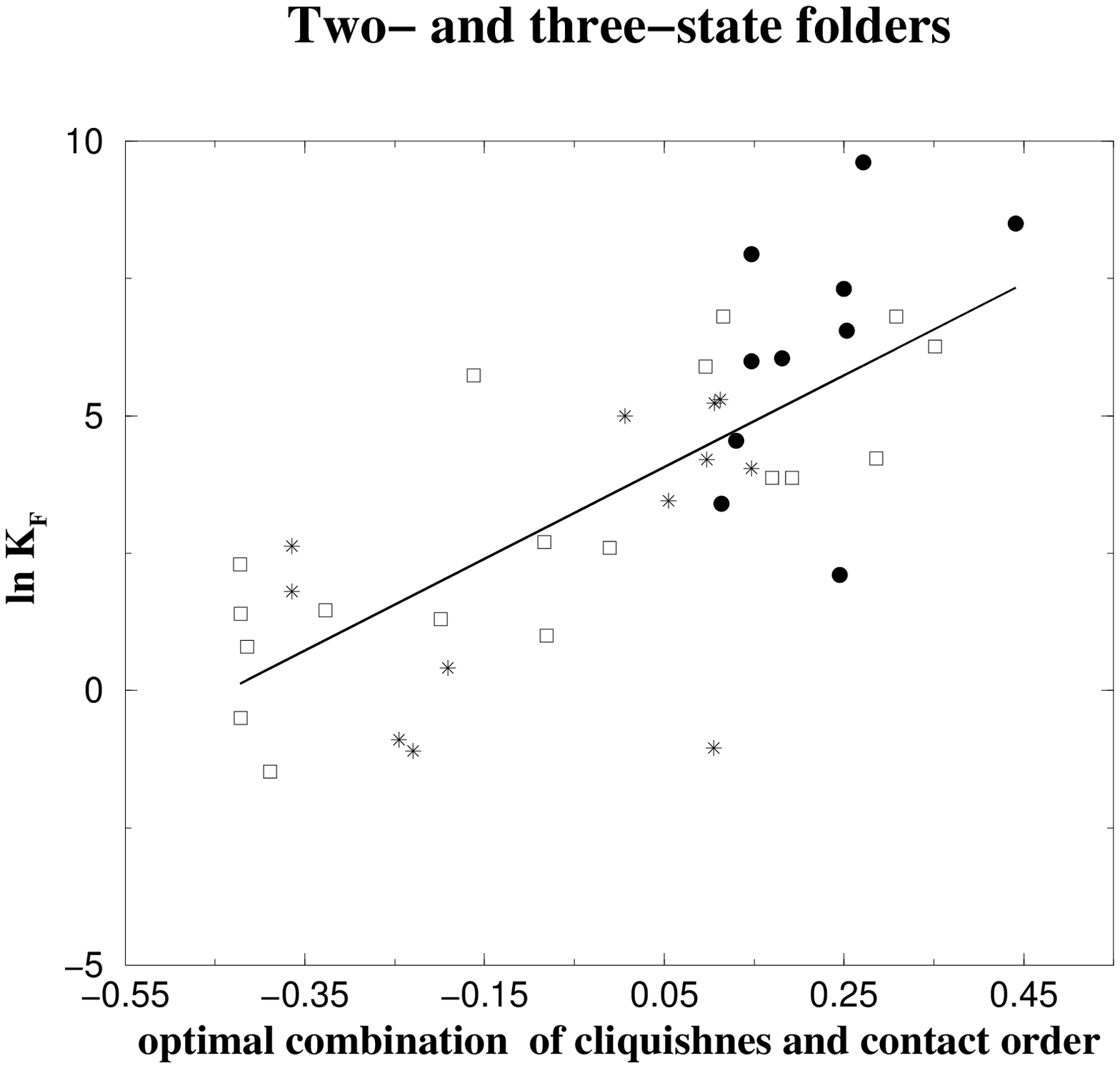,width=0.80\textwidth}
\end{center}
\caption{}
\label{fig:combined}
\end{figure}

\begin{figure}
\begin{center}
\epsfig{figure=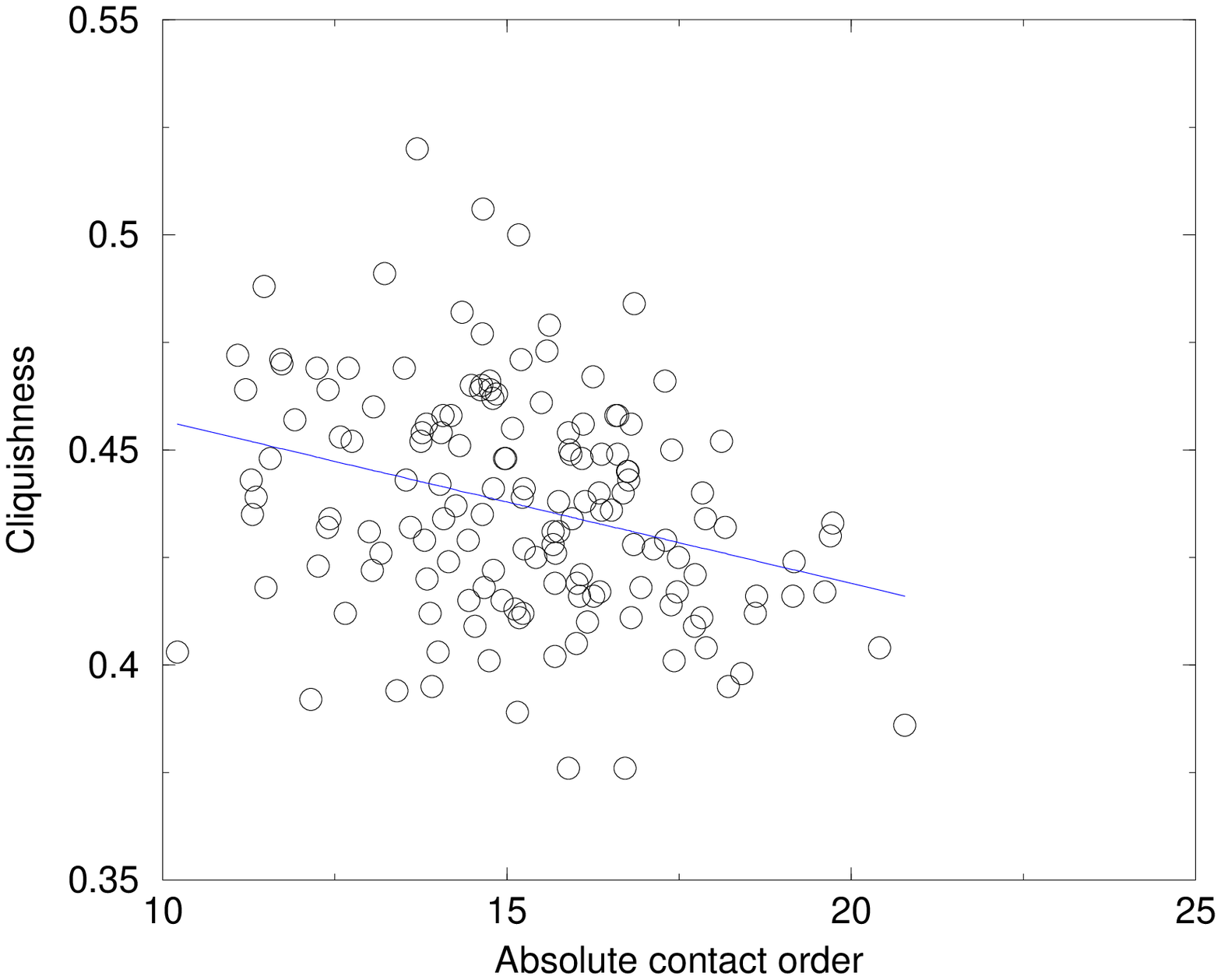,width=0.80\textwidth}
\end{center}
\caption{}
\label{fig:decoys}
\end{figure}

Nevertheless, since the mutual correlation is not perfect, it is still
possible to achieve, by definition, better predictions by combining
cliquishness and contact order. The degree of enhancement depends also
on the statistical significance of the individual starting
correlations. For these reasons, the improvement is noticeable for
folding rates, while it is not significant for transition state
placement.  For the case of two-state folders, the optimal combination
yields correlations of $r=0.86$ while for the more general case of two
and three-state folding rates one has $r=0.74$ which leads to a
discernible improvement over previous cases, as visible in Fig.
\ref{fig:combined}. To the best of our knowledge, this is the highest
correlation recorder among similar studies involving a comparable
number of entries (also including non-linear prediction schemes
\cite{karplusnsb}).  Due to the fact that the optimal combined
correlations are found {\em a posteriori}, the associated values of
$t$ are no more meaningful indicators of statistical significance.

Besides the cliquishness, we have investigated other parameters that
are routinely used to characterise general networks (networks of
contacts in our case). In particular, we considered the ``diameter''
of the contact map, defined as the largest degree of separation
between any two residues, and also its average value. The diameter
measures the maximum number of contact that need to be traversed to
connect an arbitrary pair of distinct residues. Although the
contact-map diameter is an abstract object, it conveys relevant
topological information about protein structure, since it measures the
long-range structural organisation. We found, a posteriori, that
neither the maximum, nor the average diameter, correlate in a
significant manner with the folding rate or transition state
placement.

\section{Conclusions}

We have analysed important topological descriptors of organised
networks (in our case the spatial network of native contacts) that
could be used, individually, or in mutual conjunction, to describe and
predict experimental parameters used to characterize the folding
process. It is found that, besides the previously introduced contact
order, a topologic parameter, termed cliquishness or clustering
coefficient, is a powerful indicator of both the folding velocity and
the transition state placement for two- and three-state folders. The
predicting power of the cluquishness is that it takes into account the
presence and organisation of clusters of interdependent contacts that
are putatively responsible for the cooperative formation of
native-like regions. This property appears well-suited to reproduce
important features in the transition state that are otherwise elusive
to other topologic analysis. The high statistical significance of the
observed correlations testifies the strong influence of geometric
structural issues on the folding process.  The maximum predicting
power is obtained when the topologic information contained in the
cliquishness is used in combination with the contact order; this
allows to reach a linear correlation as high as 0.74 with experimental
folding rates recorder in 40 experimental measurements.

\section{Methods}

\subsection{Cross correlations}

The linear correlation between two sets of data, $\{x\}$ and $\{y \}$
is obtained from the normalised scalar products of the covariations:

\begin{equation}
r = {\sum_i (x_i - \bar{x}) (y_i - \bar{y}) \over \sqrt{\sum_i (x_i -
\bar{x})^2} \sqrt{\sum_j (y_j - \bar{y})^2 }}
\end{equation}

Without loss of generality, in the following we shall consider the sets
of data to be with zero average and with unit norm, so that the
expression of the correlation simplifies

\begin{equation}
r ={ \vec{x} \cdot \vec{y}}
\end{equation}

We now formulate the following problem. Two sets of data, $\{x\}$ and
$\{ y \}$ have linear correlation $r_x$ and $r_y$ respectively with a
third (reference set), $\{z\}$. What is the maximum and minimum
correlations we can expect between sets $\{x\}$ and $\{ y \}$? We
assume that $r_x$ and $r_y$ are positive since this condition can
always be met by changing sign, if necessary, to the vector
components.

The answer is easily found by decomposing $\{x\}$ and $\{ y \}$ into
their components parallel and orthogonal to $\{z\}$:

\begin{equation}
\vec{x} \cdot \vec{y} = b_\parallel \ c_\parallel + b_\perp \ c_\perp
\end{equation}

\noindent Since $_\parallel \ c_\parallel$ is equal to $r_x \ r_y$,
and hence is fixed, the maximum [minimum] correlation is found when $
b_\perp$ and $ c_\perp$ are [anti]parallel. Thus,

\begin{equation}
r_x \ r_y -  \sqrt{(1 - r_x^2)(1 - r_y^2)} \le r \le
r_x \ r_y +  \sqrt{(1 - r_x^2)(1 - r_y^2)}
\end{equation}

Now we turn to a different, but related problem. How can we combine
linearly $\{x\}$ and $\{ y \}$, so to have the maximum correlation
with $\{z\}$. The generic linear combination, 
\begin{equation}
\vec{k} = { \vec{x} + b \vec{y} \over \sqrt{1 + b^2 + 2 b\ \vec{x} \cdot
\vec{y} }}
\end{equation}

\noindent leads to the following correlations

\begin{equation}
\vec{k} \cdot \vec{z} = { \  r_x + b \ \  r_y \over 
\sqrt{1 + b^2 + 2 b\ \vec{x} \cdot \vec{y} } } 
\end{equation}

\noindent The maximum is achieved for

\begin{equation}
b = { \  r_y - \vec{x} \cdot \vec{y} \  r_x 
\over r_x \ - \ \  \vec{x} \cdot \vec{y} \ \  r_y   }
\end{equation}

\noindent which yields
\begin{equation}
Max(\vec{k} \cdot \vec{z}) =
\sqrt{r_x^2 + r_y^2 - 2 \vec{x} \cdot \vec{y}
\ r_x\ r_y  
\over 1 - (\vec{x} \cdot \vec{y})^2 }
\end{equation}

\subsection{Generation of alternative compact structures}

To generate the thirty randomly-collapsed structures used in the
comparison of Fig. \ref{fig:decoys}, we adopted a Monte Carlo
technique. The length of the artificial proteins ranged uniformly in
the interval 80-110. Starting from an open conformation, each
structure was modified under the action of typical MC moves
(single-bead, crankshaft, pivot) \cite{Sokal}. A newly generated
modified configuration is accepted according to the ordinary
Metropolis rule. The energy scoring function is composed of two terms:
The first one contains a homopolymeric part that rewards the
establishment of attractive interactions (cutoff of 6.0 \AA) between
any pair of non-consecutive residues. The second term is introduced to
penalise structure realisations with radii of gyration larger than
that found in naturally-occurring proteins with the same length. The
Monte Carlo evaluation is embedded in a simulated annealing scheme
\cite{kirk} which allows to minimise efficiently the scoring function
by slowly decreasing a temperature-like control parameter.

\section{Acknowledgements}

We are indebted with Amos Maritan for several illuminating discussions
and with Fabio Cecconi and Alessandro Flammini for a careful reading
of the manuscript. Support from INFM and MURST Cofin99 is
acknowledged.


\newpage
\vfill
\begin{table}
\begin{center}
\begin{tabular}{|l|l|l|l|l|l|}
Protein & Length & Family & ln $K_f$ & $\theta_m$ & Cliquishness\\
& & & & &($d=4.6$ \AA)\\
\hline \hline
1shg \protect{\cite{1shg}} & 57 & $\alpha$ & 2.10& 0.69& 0.546\\
1lmb \protect{\cite{lmb,lmb2}} & 80 & $\alpha$ & 8.50& 0.46& 0.555\\
2abd \protect{\cite{2abd,2abd2}} & 86 & $\alpha$ & 6.55& 0.61& 0.550\\
1imq \protect{\cite{goddard}} & 86 & $\alpha$ & 7.31& --- & 0.545\\
1ycc \protect{\cite{1ycc}} & 103 & $\alpha$ & 9.61& 0.34& 0.548\\
1hrc \protect{\cite{1hrc}} & 104 & $\alpha$ & 7.94& 0.47& 0.538\\
1hrc, horse, oxidized Fe$^{III}$ \protect{\cite{1hrc}} & 104 & $\alpha$ & 5.99& 0.40& 0.538\\
2gb1 \protect{\cite{2gb1}} & 56 & $\alpha/\beta$ & 6.26& --- & 0.553\\
2ptl \protect{\cite{2ptl}} & 61 & $\alpha/\beta$ & 4.22& 0.75& 0.551\\
2ci2 \protect{\cite{1coa}} & 65 & $\alpha/\beta$ & 3.87& 0.61& 0.535\\
1cis \protect{\cite{1cis}} & 66 & $\alpha/\beta$ & 3.87& 0.61& 0.540\\
1hdn \protect{\cite{1hdn}} & 85 & $\alpha/\beta$ & 2.70& 0.64& 0.526\\
1aye \protect{\cite{1pba}} & 92 & $\alpha/\beta$ & 6.80& 0.74& 0.554\\
1urn \protect{\cite{1urn}} & 96 & $\alpha/\beta$ & 5.73& 0.55& 0.520\\
1aps \protect{\cite{1aps}} & 98 & $\alpha/\beta$ & -1.47& 0.79& 0.511\\
1fkb \protect{\cite{Jackson}} & 107 & $\alpha/\beta$ & 1.46& 0.67& 0.512\\
2vik \protect{\cite{Jackson}} & 126 & $\alpha/\beta$ & 6.80& 0.73& 0.546\\
1srl \protect{\cite{1srl}} & 56 & $\beta$ & 4.04& 0.69& 0.536\\
1shf.a \protect{\cite{1shf}} & 59 & $\beta$ & 4.55& 0.68& 0.534\\
1tud \protect{\cite{goddard}} & 60 & $\beta$ & 3.45& --- & 0.531\\
1csp \protect{\cite{1csp}} & 67 & $\beta$ & 6.04& 0.85& 0.539\\
1mjc \protect{\cite{1mjc}} & 69 & $\beta$ & 5.23& 0.91& 0.532\\
3mef \protect{\cite{1mjc}} & 69 & $\beta$ & 5.30& 0.94& 0.531\\
2ait \protect{\cite{2ait}} & 74 & $\beta$ & 4.20& 0.65& 0.542\\
1pks \protect{\cite{1pks}} & 76 & $\beta$ & -1.05& 0.60& 0.545\\
1ten \protect{\cite{1fnf10}} & 89 & $\beta$ & -1.10& 0.76& 0.511\\
1fnf, 9FN-III \protect{\cite{Igfolds}} & 90 & $\beta$ & -0.90& 0.63& 0.513\\
1wit \protect{\cite{Igfolds}} & 93 & $\beta$ & 0.41& 0.70& 0.526\\
1fnf, 10FN-III \protect{\cite{Igfolds}} & 94 & $\beta$ & 5.00& 0.65& 0.536
\end{tabular}
\end{center}
\caption{List of proteins known to fold via a two-state mechanism. 
The experimental quantities $K_F$ (s$^{-1}$) and $\theta_m$ are
desumed from the cited references. The reported cliquishness values
are calculated for the cutoff $d=4.6$ \AA yielding optimal
correlations against folding rates.}
\label{tab:list1}
\end{table}

\begin{table}
\begin{center}
\begin{tabular}{|l|l|l|l|l|l|}
Protein & Length & Family & ln $K_f$ & $\theta_m$ & Cliquishness\\
& & & & &($d=4.6$ \AA)\\
\hline \hline
1bta  & 89  & $\alpha$                       &   3.40  & 0.87  &  0.532\\
1ubq  & 76  &$\alpha/\beta$                   &   5.90  & 0.59  &  0.532\\
1bni  & 108 & $\alpha/\beta$                  &   2.60  & 0.88  &  0.524\\
1hel  & 129 & $\alpha/\beta$                  &   1.30  & 0.75  &  0.507\\
3chy & 128 & $\alpha/\beta$  & 1.0 & 0.71 & 0.512\\
1dk7  & 146  & $\alpha/\beta$                 &   0.80  & 0.78  &  0.513\\
2rn2, Urea, pH 5.5  & 155 & $\alpha/\beta$    &  -0.50  & 0.80  &  0.502\\
2rn2, GdnHCl, pH 5.5  & 155  & $\alpha/\beta$ &   1.40  & 0.63  &  0.502\\                      
1php.n  & 175 & $\alpha/\beta$                &   2.30  & 0.84  &  0.505\\
1hng, pH 7.0  & 97  & $\beta$                &   1.80  & 0.68  &  0.502\\
1hng, pH 4.5  & 97  & $\beta$                &   2.63  & 0.62  &  0.502\\
\end{tabular}
\end{center}
\caption{List of proteins known to fold via a three-state mechanism.
The experimental quantities $K_F$ (s$^{-1}$) and $\theta_m$ are
desumed from \protect{\cite{Jackson}}.  The reported cliquishness
values are calculated for the cutoff yielding optimal
correlations against folding rates.}
\label{tab:list2}
\end{table}

$\null$
\newpage

\centerline{\LARGE Figure captions}
\begin{itemize}

\item{Fig. 1. Correlation of cliquishness, relative and absolute
contact order against folding rates of two-state folders. The values
of the correlation coefficients are plotted as a function of the
cutoff, $d$, used in the definition of the contact map.}

\item{Fig. 2. Correlation of cliquishness, relative and absolute
contact order against transition state placement of two-state
folders. The values of the correlation coefficients are plotted as a
function of the interaction cutoff, $d$.}

\item{Fig. 3. Correlation of cliquishness, relative and absolute
contact order against folding rates of two- and three-state folders. The values
of the correlation coefficients are plotted as a function of the
cutoff, $d$, used in the definition of the contact map.}

\item{Fig. 4. Correlation of cliquishness, relative and absolute
contact order against transition state placement of two- and three-state
folders. The values of the correlation coefficients are plotted as a
function of the interaction cutoff, $d$.}

\item{Fig. 5. Scatter plot of cliquishness (left) and absolute contact
order (right) versus folding rates of the 40 entries of Tables 1 and
2. The used values of $d$ are the optimal ones reported in the text.
Filled circles, open squares and starred points denote proteins
belonging to the $\alpha$, $\alpha/\beta$ and $\beta$ families,
respectively.}

\item{Fig. 6. Scatter plot of cliquishness (left) and absolute contact
order (right) versus $\theta_m$ of the entries of Tables 1 and
2. The used values of $d$ are the optimal ones reported in the text.
Filled circles, open squares and starred points denote proteins
belonging to the $\alpha$, $\alpha/\beta$ and $\beta$ families,
respectively.}

\item{Fig. 7. Scatter plot of the logarithm of folding rates for the
entries of Tables 1 and 2, against data from optimally combined
cliquishness and contact order. The optimal linear superposition, see
Methods, is obtained for $b=0.7$, ($\{x\}$ and $\{y\}$ being the
cliquishness and contact order data respectively. Filled circles, open
squares and starred points denote proteins belonging to the $\alpha$,
$\alpha/\beta$ and $\beta$ families, respectively.}

\item{Fig. 8. Scatter plot of average cliquishness versus absolute
contact order, for randomly collapsed structures generated by
stochastic numerical methods.}

\end{itemize}

\newpage

\newpage

\newpage

\end{document}